\documentclass[final,5p,times,twocolumn,authoryear]{elsarticle}

\usepackage{amsmath,amsthm,amsfonts}
\usepackage{amssymb}

\usepackage[mathscr]{eucal} 
\usepackage{graphicx}

\usepackage{natbib}
\newcommand{\flow}[1]{\mathscr{#1}}
\newcommand{\vect}[1]{\boldsymbol {#1}}
\newcommand{\tens}[1]{\boldsymbol{\mathsf{#1}}}
\newcommand{\vecti}{\boldsymbol{{I}}}

\newcommand{\tscale}{\theta}
\newcommand{\Vscale}{\mathcal{V}}

\allowdisplaybreaks[4]
\begin{document}

\begin{frontmatter} 

\title{{Minimisation of a Free-Energy-Like Potential for Non-Equilibrium Systems at Steady State}}
\author{Robert K. Niven}
\ead{r.niven@adfa.edu.au}
\address{School of {Engineering and Information Technology}, The University of New South Wales at ADFA, Canberra, ACT, 2600, Australia.} 

\date{28 February 2009; revised 10 August 2009}

\begin{abstract}

This study examines a new formulation of non-equilibrium thermodynamics, which gives a conditional derivation of the ``maximum entropy production'' (MEP) principle for flow and/or chemical reaction systems at steady state. The analysis uses a dimensionless potential function $\phi_{st}$ {for non-equilibrium systems}, analogous to the free energy concept of equilibrium thermodynamics. Spontaneous reductions in $\phi_{st}$ arise from increases in the ``flux entropy'' of the system - a measure of the variability of the fluxes - or in the local entropy production; {conditionally, depending on the behaviour of the flux entropy, the formulation reduces to the MEP principle.  The inferred steady state is also shown to exhibit high variability in its instantaneous fluxes and rates, consistent with the observed behaviour of turbulent fluid flow, heat convection and biological systems; one consequence is the coexistence of energy producers and consumers in ecological systems. The different paths for attaining steady state are also classified.} 

\end{abstract}


\end{frontmatter} 


%
%


\section{\label{Intro}Introduction} 
%

Since the seminal book ``What is Life" by Erwin \citet{Schrodinger_1944}, scientists have pondered the existence of life and its compatibility with the second law of thermodynamics. Ridiculing the popular notion that the primary purpose of {biological metabolism} is to extract matter and energy from the environment, he moves to the crux of the issue (p 71):
\begin{quote}
``{\it ... a living organism continually increases its entropy ... and thus tends to approach the dangerous state of maximum entropy, which is death. It can only keep aloof from it, i.e.\ alive, by continually drawing from its environment negative entropy ... What an organism feeds upon is negative entropy.}''
\end{quote}
(The argument is qualified in a footnote, to refer to free energy instead of negative entropy.) The topic was taken up in more detail by \citet{Prigogine_1967, Prigogine_1980}, 
who described living organisms {\textendash } along with heat-transporting convection cells, turbulent fluid flow vortices and oscillatory chemical reactions {\textendash } as {\it dissipative structures}, which continually dissipate heat and thus generate and export entropy to the environment. However, Prigogine's main quantitative result, his {\it minimum entropy production} (MinEP) principle {\textendash } valid in the linear or \citet{Onsager_1931a, Onsager_1931b} transport regime {\textendash } seems diametrically opposed to life {\citep{Martyushev_etal_2007}, as was recognised by \citet[][p88]{Prigogine_1980} himself}. Bacteria in a microcosm, organisms in an ecosystem or humans on a planet do not try to minimise their entropy production, but instead grow, reproduce and consume all available resources as rapidly as possible. More recently, other thermodynamics-inspired perspectives on biological systems have been advanced, including the use of biological measures of entropy and information \citep[e.g.][]{Ayres_1994}; the non-mathematical gradient theory of \citet{Schneider_S_2005}; and exergy-based treatments of ecological systems and processes \citep[e.g.][]{Jorgensen_2006}. 

Over the past 30 years, a new principle has {been proposed}, the {\it maximum entropy production} (MEP) principle, which states that a flow system subject to various flows or gradients will tend towards a steady state position of maximum thermodynamic entropy production, $\dot{\sigma}$ \citep{Ozawa_etal_2003, Kleidon_L_book_2005, Martyushev_S_2006, Bruers_2007c}. The MEP principle has been successfully applied {\textendash } in a heuristic sense {\textendash } to the prediction of steady states of a wide range of systems, including the Earth's climate system \citep[e.g.][]{Paltridge_1975, Paltridge_1978, Kleidon_2004, Kleidon_L_book_2005}; thermal (B\'enard) 
convection \citep{Ozawa_etal_2001}; mantle convection \citep{Vanyo_Paltridge_1981, Lorenz_2001b}; electrical currents \citep{Zupanovic_etal_2004, Botric_etal_2005, Christen_2006, Bruers_etal_2007a}; plasmas \citep{Christen_2007a, Yoshida_M_2008}; crystalline solids \citep{Martyushev_A_2003, Christen_2007b}; ecological systems \citep{Meysman_B_2007} and biochemical processes \citep{Juretic_Z_2003, Dewar_etal_2006}. The MEP principle therefore {offers a new approach} for the analysis of biological systems at the cellular, organism, ecosystem and biosphere levels. Most importantly, {it is a {\it quantitative}} principle, based on precisely defined, rigorous thermodynamic concepts; {it} does not rest upon vague, non-mathematical notions such as ``order'', ``disorder'', ``randomness'' or ``complexity'' often seen in discussions of biological systems.

Several theoretical justifications of the MEP principle have been advanced, including approaches based on path or transition probabilities \citep{Dewar_2003, Dewar_2005, Attard_2006a, Attard_2006b} and two more generalistic arguments \citep{Zupanovic_etal_2006, Martyushev_2007}. Recently, a rather different derivation was presented {to {\it directly} determine the steady state of a flow system}, based on an entropy defined on the set of local instantaneous fluxes and reaction rates through or within {each infinitesimal element; this reduces to} a local form of the MEP principle in some circumstances \citep{Niven_MEP}.  The analysis invokes a generalised potential function (negative Massieu function) {obtained from Jaynes' maximum entropy method}, somewhat analogous to the free energy concept used in equilibrium thermodynamics, which attains a minimum at steady state. The {{\it aim}} of this study is to explore the implications of this derivation in somewhat simpler terms than in \citet{Niven_MEP}, using terminology adapted from chemical and statistical thermodynamics. {In particular, the nature of the inferred steady state of a flow system, and the various means by which it can be attained, are examined in detail.} The analysis has important implications for the modelling of flow systems, {including} the Earth's climatic-biosphere system and all biological systems.
%
\section{\label{Gen}The Generalised Free Energy Concept} 
\subsection{\label{Theory}Jaynes' MaxEnt}
%
The maximum entropy (MaxEnt) principle of \citet[][see also \citealp{Tribus_1961a, Tribus_1961b, Kapur_K_1992}]{Jaynes_1957, Jaynes_1963, Jaynes_2003} 
provides a powerful technique with which to infer the most probable position of a probabilistic system. Consider a system composed of $N$ distinguishable entities allocated to $s$ equiprobable, distinguishable categories (a {\it multinomial system}: \citealp{Niven_2005, Niven_2006, Niven_2007, Niven_2009, Niven_G_2009}). In the asymptotic limit $N\to\infty$, the most probable position can be obtained by maximising the {relative entropy function (the negative of the Kullback-Leibler \cite{Kullback_L_1951} function, $D$):}
\begin{equation}
{
\mathfrak{H} = -D = - \sum\limits_{i = 1}^s {p_i \ln \frac{{p_i }}{{q_i }}} 
}
\label{eq:relent}
\end{equation}
where $p_i$ is the probability of an entity in the $i$th category {and $q_i$ is the source or ``prior'' probability of category $i$. For equiprobable categories, this reduces to the \citet{Shannon_1948} entropy function:}
\begin{equation}
\mathfrak{H}_{Sh} =   - \sum\limits_{i = 1}^s {p_i \ln p_i } 
\label{eq:Shannon}
\end{equation}
{plus a constant. Eq.\ \eqref{eq:relent} (or \eqref{eq:Shannon}) is maximised} subject to the natural {(normalisation)} and any moment constraints on the system:
\begin{gather}
\sum\limits_{i = 1}^s {p_i }= 1,
\label{eq:C0} 
\\
\sum\limits_{i = 1}^s {p_i f_{ri} }= \left\langle {f_r } \right\rangle, \quad r = 1,...,R, 
\label{eq:Cr}
\end{gather}
where $f_{ri}$ is the value of the $i$th category of property $f_r$  and $\left\langle {f_r} \right\rangle$ is the expectation (average) of $f_{ri}$. This yields the most probable (stationary) distribution of the system:
\begin{gather}
\begin{split}
p_i^* &=  q_i \exp \Bigl({  - \lambda_0^*   - \sum\limits_{r = 1}^R  {\lambda_r  f_{ri} }    }\Bigr) 
=  (Z^*)^{-1} q_i { \exp \Bigl({  - \sum\limits_{r = 1}^R  \lambda_r  f_{ri}   } \Bigr) }  , \\
Z^* &=  q_i \exp({\lambda_0^*}) = \sum\limits_{i=1}^s {\exp\Bigl( {  - \sum\limits_{r = 1}^R  \lambda_r  f_{ri}   } \Bigr)} ,
\end{split}
\label{eq:pstar2_i}
\end{gather}
and the maximum entropy position \citep{Jaynes_1957, Jaynes_1963, Jaynes_2003}:
\begin{gather}
{ \mathfrak{H}^*  = \lambda_0^*  +  \sum\limits_{r = 1}^R \lambda_r \left\langle {f_r} \right\rangle },
\label{eq:Hstar}
\end{gather}
where $\lambda_r$ is the $r$th Lagrangian multiplier, $\lambda_0^*$ is the ``Massieu function'' \citep{Massieu_1869}, $Z^*$ is the partition function and an asterisk denotes the stationary position. It is emphasised that the above derivation is generic, and applies to {\it any} probabilistic system of multinomial form; it need not refer to a thermodynamic system.

\subsection{\label{Potential}{Generalised Heat, Work and Potential Function}}

{We now consider any} conserved quantity $f_r$, {for which we} adopt the definition:
\begin{gather}
d\langle f_r \rangle = \delta W_r + \delta Q_r 
\label{eq:work_heat}
\end{gather}
{where the path differentials $\delta W_r= \sum\nolimits_{i=1}^s  p_i^* d f_{ri} $ and $\delta Q_r=\sum\nolimits_{i=1}^s  d p_i^* f_{ri}$ can be termed the ``generalised work'' and ``generalised heat'' associated} {respectively} with a change in $\langle f_r \rangle$. {It} can be shown \citep{Jaynes_1957, Jaynes_1963, Jaynes_2003} that:
\begin{gather}
{d\mathfrak{H}^* 
= \sum\limits_{r = 1}^R {\lambda _r \delta Q_r } }
\label{eq:dHstar}
\end{gather}
This is a ``generalised Clausius equality'' \citep[c.f.][]{Clausius_1865}, applicable to all multinomial systems. Substituting \eqref{eq:dHstar} into the differential of \eqref{eq:Hstar} and rearranging gives:
\begin{align}
\begin{split}
d \phi &= - d \lambda_0^*  =  \sum\limits_{r = 1}^R {\lambda _r  \delta W_r }  +  \sum\limits_{r = 1}^R d \lambda_r \langle {f_r} \rangle
\\
&=-d\mathfrak{H}^*  +  d \, \Biggl( \sum\limits_{r = 1}^R \lambda_r \langle {f_r} \rangle \Biggr)
\end{split}
\label{eq:Massieu}
\end{align}
We therefore obtain a potential function $\phi$ (negative Massieu function) which captures all possible changes in the system, due to changes in the entropy $\mathfrak{H}^*$ or in the ``constraint set'' $\sum\nolimits_{r = 1}^R \lambda_r \left\langle {f_r} \right\rangle$. If the multipliers $\{\lambda_r\}$ are constant, $d\phi$ reduces to the multiplier-weighted total generalised work on the system, $\sum\nolimits_{r = 1}^R {\lambda _r  \delta W_r }$. We therefore see that $\phi$ is a dimensionless, weighted, extended version of the free energy function, applicable to {\it any} probabilistic system of multinomial form \citep{Jaynes_1957, Jaynes_1963, Jaynes_2003, Tribus_1961a, Tribus_1961b}.

How should we interpret \eqref{eq:Massieu}?  {Consider some form of ``open system'', consisting of a defined region or collection of discrete entities in contact with some surroundings (or the rest of the universe). The internal structure of the system may be described by some probability function $p_i$, giving rise to some relative entropy function $\mathfrak{H}$ for the system (not necessarily the thermodynamic entropy $S$, but {\it any} entropy). From a purely probabilistic formulation of the second law \citep{Niven_2009, Niven_MEP}:
\begin{quote}
``{\it The entropy of the universe, $\mathfrak{H}_{univ}$, however defined, can only increase''},
\end{quote}
it is evident that any spontaneous event must result in an increase of the entropy of the system, $\mathfrak{H}^*$, and/or an increase in entropy produced and exported by the system to its surroundings, $\mathfrak{H}_{prod}$.} {Quantitatively, this can be written as\footnote{{Technically, the variation in $\mathfrak{H}_{prod}$ is written with a $\delta$, since it is a ``non-property'' of the system; however, for a reproducible phenomenon, it will be expressible in terms of other state functions of the system.}}:
\begin{gather}
d \mathfrak{H}_{univ} =d\mathfrak{H}^* +  \delta \mathfrak{H}_{prod} \ge 0
\label{eq:H_univ}
\end{gather}}
{However, the only means by which a system can produce and export entropy -- thereby increasing $\delta \mathfrak{H}_{prod}$ -- is by a reduction in the magnitude of one or more constraints (or multipliers) which govern the system, $\{\langle f_r \rangle \}$ (or $\{ \lambda_r \}$).  For such a change to give rise to a quantity in (dimensionless) entropy units, we therefore establish that $\delta \mathfrak{H}_{prod} =- d (\sum\nolimits_{r = 1}^R \lambda_r \langle {f_r} \rangle)$. Comparing \eqref{eq:Massieu} and \eqref{eq:H_univ}, we thus see that $d \phi$ expresses, in a negative sense, the change in entropy of the universe. This can be written as:
\begin{gather}
d \phi =-d\mathfrak{H}^*  -  \delta \mathfrak{H}_{prod} \le 0
\label{eq:Massieu2}
\end{gather}
Eqs.\ \eqref{eq:Massieu} and \eqref{eq:Massieu2} thus provide a mathematical formulation of a generalised second law (with sign reversed), expressing the interplay between changes in the entropy -- however defined -- of a system, and changes in entropy produced and exported by a system to its surroundings. This is again consistent with the interpretation of $\phi$ as a dimensionless, weighted, extended version of the free energy concept \citep{Jaynes_1957, Jaynes_1963, Jaynes_2003, Tribus_1961a, Tribus_1961b}.
}

\section{\label{Apps}{Applications}} 
\subsection{\label{Eg1}Equilibrium Systems Example}

The above discussion is best illustrated by an example from thermodynamics. Whilst a broader free energy concept is considered in \citet{Niven_MEP}, most readers will be more familiar with the \citet{Gibbs_1875} free energy function for systems of constant composition:
\begin{equation}
G=-TS^* +  U  + PV = -TS^* + H, 
\label{eq:Gibbs}
\end{equation}
where $S^*$ is the maximum thermodynamic entropy, $U$ is internal energy, $V$ is volume, $P$ is pressure, $T$ is absolute temperature and $H$ is the enthalpy. Eq.\ \eqref{eq:Gibbs} can be derived by applying Jaynes' method to an equilibrium thermodynamic system subject to the constraints $U$ and $V$, wherein $p_i$ is the joint probability that a molecule will occupy a specified energy level and volume element. Eq.\ \eqref{eq:Massieu} then gives the potential function \citep{Jaynes_1957, Jaynes_1963, Tribus_1961a, Tribus_1961b, Jaynes_2003}:
\begin{gather}
\begin{split}
d \phi_{eq} =-d\mathfrak{H}^*_{eq}  +  d \bigl(\lambda_U  U  + \lambda_V  V  \bigr)
\end{split}
\label{eq:Massieu_eq}
\end{gather}
Recognising $S^*=k \mathfrak{H}^*_{eq}$, $\lambda_U=1/kT$ and $\lambda_V=P/kT$, where $k$ is Boltzmann's constant, gives:
\begin{gather}
k d \phi_{eq} = d \Bigl( \frac{G}{T} \Bigr) = -dS^*  +  d \Bigl( \frac{H}{T} \Bigr) \le 0
\label{eq:Massieu_eq2}
\end{gather}
equivalent to \eqref{eq:Gibbs}. This form reveals the {true} meaning of the {Gibbs} free energy concept: it expresses {\textendash } in a negative sense {\textendash } the interplay between the change in entropy {within} the system, $dS^*$, and the change in entropy {exported by the system}, $-d(H/T)$, due to transfers of heat \citep{Planck_1922, Planck_1932, Fermi_1936, Strong_H_1970, Craig_1988}. From the (classical) second law, their sum must be positive, and so a system will spontaneously approach a position of minimum $G/T$ (for constant $T$, it will approach minimum $G$). From \eqref{eq:Massieu2}, we can rewrite \eqref{eq:Massieu_eq2} as \citep{Niven_MEP}:
\begin{gather}
k d \phi_{eq} = d \Bigl( \frac{G}{T} \Bigr) = -dS^*  - \delta  \sigma \le 0
\label{eq:Massieu_eq3}
\end{gather}
where {$\delta \sigma = k \delta \mathfrak{H}_{prod}$} is the increment of {\it thermodynamic entropy produced {and exported}} {by the system to its surroundings.}

\subsection{\label{Eg2}Flow System Example}

Now consider a second example, of an infinitesimal fluid element in a control volume of a flowing fluid, subject to local mean values of the heat flux ${\vect j}_Q$, diffusive mass fluxes ${\vect j}_{c}$ of each chemical species $c$, stress tensor ${\tens{\tau}}$ and chemical reaction rates $\hat{\dot{\xi}}_d$ of each reaction $d$, plus the natural constraint (\ref{eq:C0}). This  model encompasses all biological and ecological systems. Such a system can be analysed by Jaynes' method using the local flux {relative} entropy \citep{Niven_MEP}:
\begin{equation}
{
\mathfrak{H}_{st} = - \sum\limits_{\vecti} {\pi_{\vecti} \ln \frac{{\pi_{\vecti} }}{{\gamma_{\vecti} }}} 
}
\label{eq:h_st}
\end{equation}
where $\pi_{\vecti}$ is the joint probability that the fluid element experiences a set of instantaneous local fluxes of heat, species $c$, momentum and rates of chemical reactions $d$, {and $\gamma_{\vecti}$ is the joint prior probability}. For this system, it can be shown that \eqref{eq:Massieu} gives the increment in the local potential function \citep{Niven_MEP}:
\begin{equation}
d\phi_{st} = - d\mathfrak{H}_{st}^* - \frac {\tscale \Vscale}{k} \delta \hat{\dot{\sigma}}  \le 0 
\label{eq:Massieu_st}  
\end{equation}
where $\mathfrak{H}_{st}^*$ is the local flux entropy at steady state, $\tscale$ and $\Vscale$ are characteristic time and volume scales of the system, and $\hat{\dot{\sigma}}$ is the local {\it thermodynamic entropy production} of the element per unit volume:
\begin{gather}
\begin{split}
\hat{\dot{\sigma}}= 
{\vect j}_Q \cdot {\vect \nabla} \biggl({\frac{1}{T}} \biggr) 
- \sum\limits_{c} {{\vect j}_c} \cdot \biggl[{\vect \nabla} \biggl(\frac{\mu_c}{M_c T} \biggr) - \frac {{\vect g}_c}{T}   \biggr]
\\
 -  {{\tens{\tau}}} : { \vect \nabla} \biggl(\frac{{\vect v} }{T} \biggr)^\top
 -  \sum\limits_{d} \hat{\dot{\xi}}_{d} \frac{A_{d}}{T}  
\end{split}
\label{eq:sigma_dot_hat2}
\end{gather}
in which $\mu_c$, $M_c$ and ${\vect g}_c$ are the molar chemical potential, molar mass and specific body force on species $c$, $\vect{v}$ is the mass-average fluid velocity and $A_{d}$ is the molar chemical affinity of reaction $d$, with $A_{d}<0$ indicating spontaneous forwards reaction. Eq.\ \eqref{eq:sigma_dot_hat2} can be summarised in the form:
\begin{equation}
{
\hat{\dot{\sigma}}= \sum\limits_X j_X F_X
}
\label{eq:LEP}
\end{equation}
where $j_X$ is the local mean flux or mean reaction rate of quantity or species $X$, and {$F_X$} is the corresponding local ``thermodynamic force'' (gradient or affinity term). Not coincidentally, \eqref{eq:Massieu_st} has the same form as \eqref{eq:Massieu2}, expressing (with sign reversed) the sum of changes in the flux entropy {within} the element plus {its export out} of the system. A flow element will therefore try to approach a steady state position of minimum $\phi_{st}$, for the same reason that an equilibrium thermodynamic system tries to approach an equilibrium position of minimum $\phi_{eq}$ (minimum $G/T$). 

Comparing \eqref{eq:Massieu_eq2} and \eqref{eq:Massieu_st}, we see that the entropy production $\hat{\dot{\sigma}}$ within a fluid element of a steady state system plays a similar role (with change of sign and units) to the enthalpy function $H$ in equilibrium systems. This is an important insight, which has perhaps been hindered by the lack of popular understanding of the free energy concept \citep[see][]{Strong_H_1970, Craig_1988}. The common feature is that $H$ and $\hat{\dot{\sigma}}$ both serve as (modified) measures of the export of entropy {\textendash } however defined {\textendash } by a system to its surroundings. Many previous authors have erred in considering $\hat{\dot{\sigma}}$ to be the non-equilibrium analogue of $S^*$; whilst this may seem reasonable at first glance, it is not correct. 

\section{\label{Imp}Implications}
\subsection{\label{FluxEnt}Meaning of the Flux Entropy}
\begin{figure*}[t]
\begin{center}
\setlength{\unitlength}{0.6pt}
  \begin{picture}(800,150)
   \put(540,-10){\includegraphics[width=50mm]{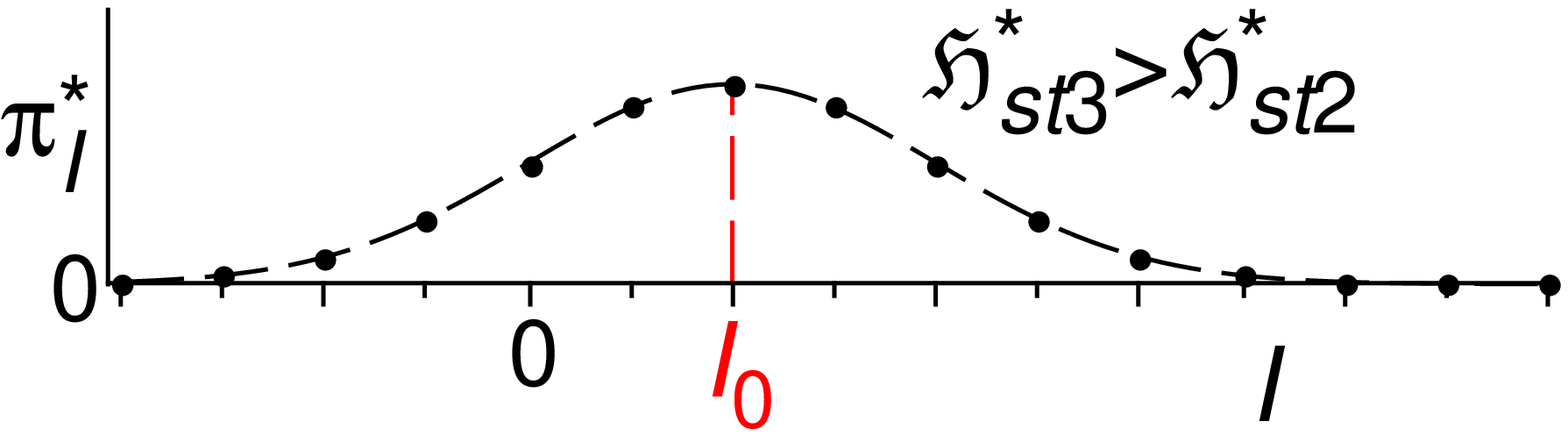} }
   \put(540,-20){(c)}
   \put(270,-10){\includegraphics[width=50mm]{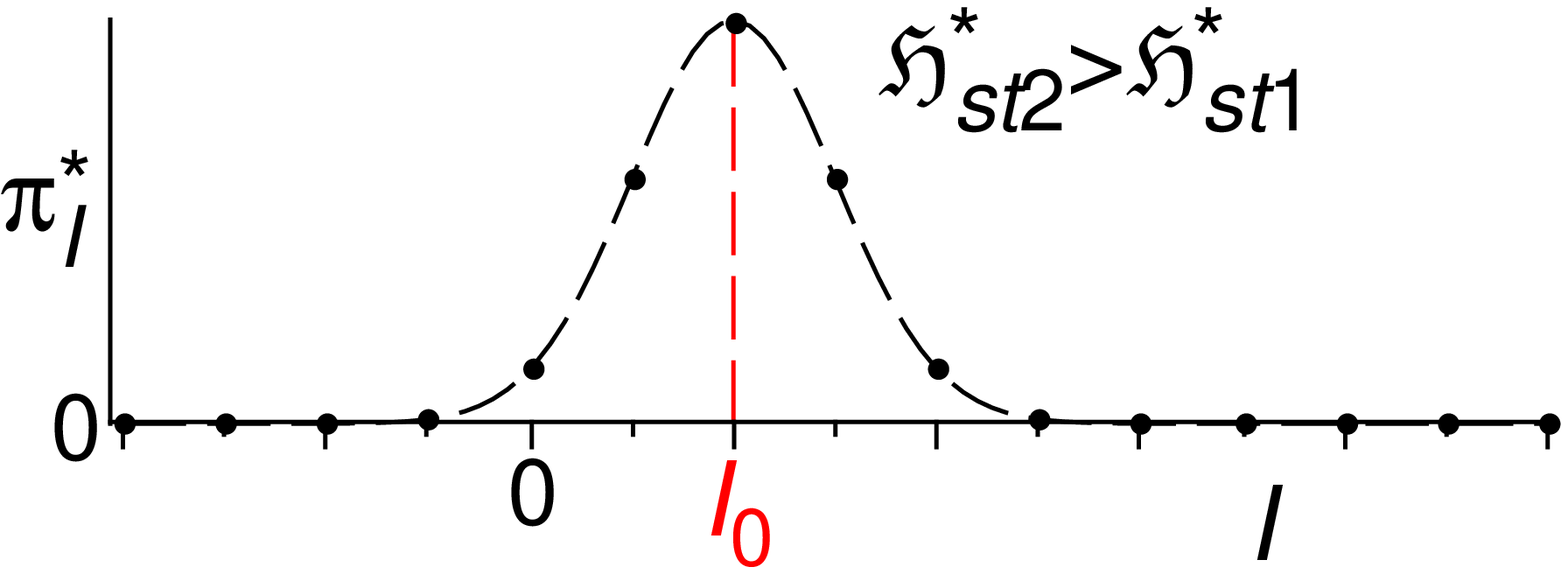} }
   \put(270,-20){(b)}
  \put(0,-10){\includegraphics[width=50mm]{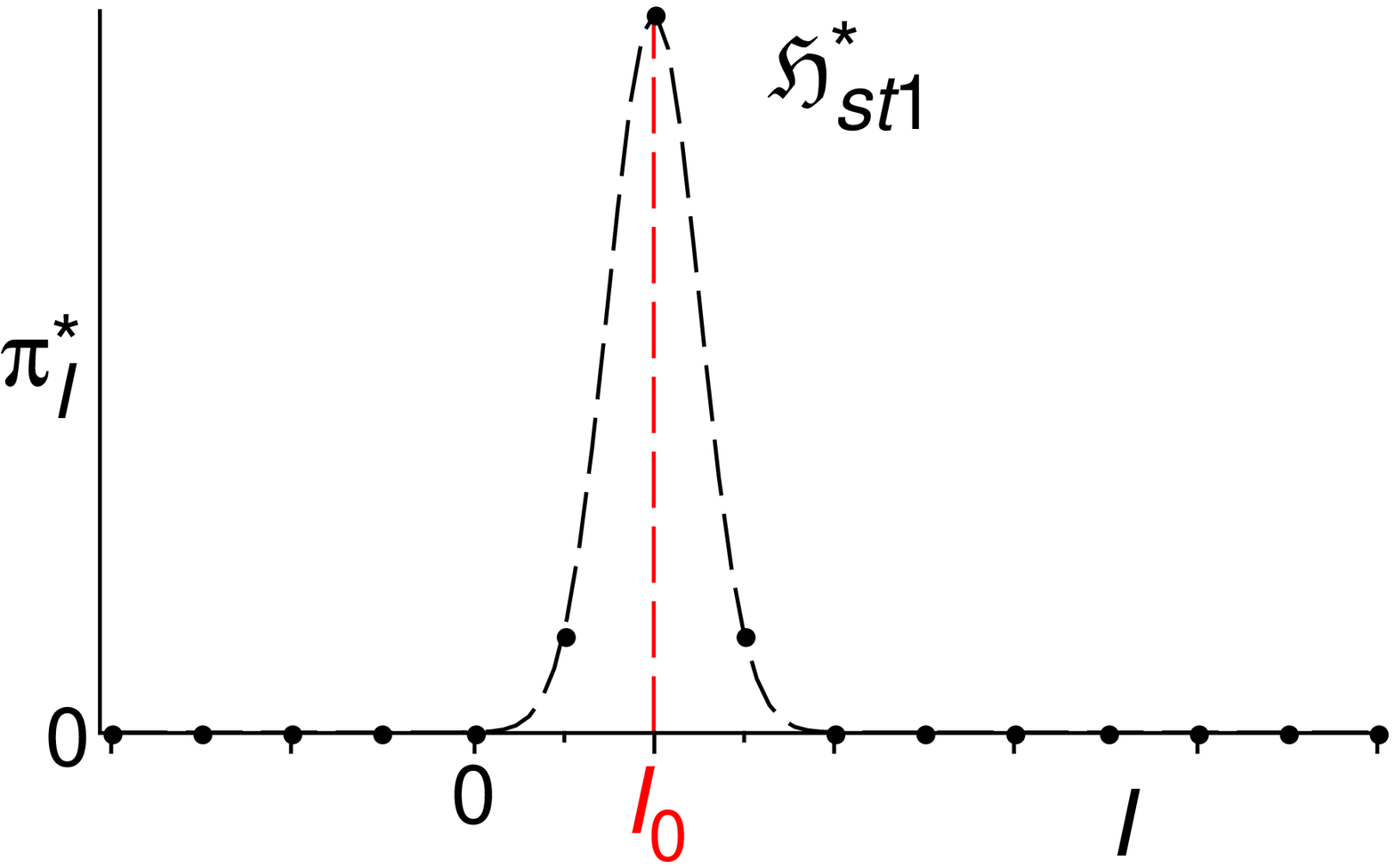} }
   \put(0,-20){(a)}
  \end{picture}
\end{center}
\caption{Effect of increasing $\mathfrak{H}_{st}^*$ on a univariate steady-state distribution $\pi_I^*$ (schematic only).} 
\label{fig:h_st}
\end{figure*}

To understand \eqref{eq:Massieu_st}, it is necessary to appreciate the meaning of the flux entropy $\mathfrak{H}_{st}^*$. To do this, we need to consider the mathematical properties of the Shannon entropy \eqref{eq:Shannon} \citep[e.g.][chap.\ 2]{Kapur_K_1992}. In essence, $\mathfrak{H}$ indicates the spread of the distribution $p_i$ amongst its categories $i$; the thermodynamic entropy $S^*$ therefore reflects the spread of the equilibrium distribution $p_i^*$ over energy levels and volume elements, with low $S^*$ indicating a narrow distribution and high $S^*$ a broad one. In the same way, $\mathfrak{H}_{st}^*$ reflects the spread of the steady state probability $\pi_{\vecti}^*$ over the set of instantaneous local fluxes and reaction rates.  This is illustrated by the schematic plots in Figures \ref{fig:h_st}a-c, for a univariate parameter $\vecti=I$ (e.g., a single flux or reaction rate of quantity $X$). All three plots have the same mean flux or rate $j_X$ {\textendash } represented schematically by a fixed ``mean category'' $I_0$ {\textendash } but the variance, and therefore the flux entropy, increases to the right. 

The plots reveal an additional, extraordinary feature of flow systems. In equilibrium systems, the categories (e.g.\ energy levels) are generally taken to start from a ``zero'' or reference level, for which the value of the index is unimportant. In contrast, flow systems have no such minimum, since we must allow for positive and negative flux or rate levels 
$I = 0, \pm1, \pm2, ...$.
In consequence, as $\mathfrak{H}_{st}^*$ increases, the system is more likely to access its states of reverse flow or reverse chemical reaction $I<0$, even if the mean value $j_X$ is high (see Figures \ref{fig:h_st}b-c). In other words, a high flux entropy is associated with {\it greater {variability}} in the fluxes and rates, which therefore implies {\it oscillatory} or {\it chaotic} processes.  We immediately see the connection between steady states of high $\mathfrak{H}_{st}^*$ and the defining features of many ``far from equilibrium'' systems, such as fluid turbulence, heat-induced convection cells, nonlinear diffusion phenomena and oscillatory chemical reactions \citep{Prigogine_1967, Prigogine_1980}; {indeed, the latter are prevalent in biochemical processes such as} nutrient degradation processes \citep{Meysman_B_2007} and the photosynthesis cycle \citep{Juretic_Z_2003, Dewar_etal_2006}. 

{The importance of above analysis can be illustrated by its application to the species population structure within an ecosystem. Consider a small element of a (rudimentary) ecosystem of $s$} {species, identified only by their energy usage, such that each organism of species $i$ has the energy consumption $\epsilon_i$, whilst the system has mean energy consumption $\langle E \rangle$. This model dramatically simplifies the MaxEnt ecosystem model given by \citet{Dewar_P_2008}. To infer the steady state population distribution $\pi_i^*$, we maximise the relative entropy $\mathfrak{H}_{ecol}=- \sum\nolimits_{i=1}^s \pi_i \ln \pi_i/\gamma_i$ \eqref{eq:h_st}, subject to known prior probabilities $\gamma_i$ and constraints $\sum\nolimits_{i=1}^s \pi_i =1$ and $\sum\nolimits_{i=1}^s \pi_i \epsilon_i = \langle E \rangle$, giving:}
\begin{equation}
\begin{split}
{
\pi_i^* =   \flow{Z}^{-1} \gamma_i \exp(- \zeta_E \epsilon_i)
}
\\
{
\flow{Z} =  e^{\zeta_0} = \sum\limits_{i=1}^s \gamma_i \exp(- \zeta_E \epsilon_i)
}
\end{split}
\label{eq:pi_ecol}
\end{equation}
{where $\zeta_0$ and $\zeta_E$ are, respectively, the Lagrangian multipliers for the two constraints, and $\flow{Z}$ is the partition function. The analysis \eqref{eq:Massieu_st}-\eqref{eq:LEP} then follows from \eqref{eq:pi_ecol}, with the energetic multiplier identified as $\zeta_E \propto F_E = \vect{\nabla} T^{-1}$. Although \eqref{eq:pi_ecol} has the appearance of a Boltzmann distribution akin to that of chemical thermodynamics, in an ecosystem the ``energy levels'' $i$ are actually ``energy consumption levels'', which can be positive} {or negative, corresponding respectively to {\it net energy consumers} ($i>0$) and {\it net energy producers} ($i<0$).  At a high ecological flux entropy $\mathfrak{H}_{ecol}^*$, the ``most probable'' ecosystem will therefore be forced to contain both energy producers and consumers, rather than just energy consumers. Similarly, in turbulent fluid flow, some energetic structures will be net energy consumers (dissipating energy as heat), whilst others will be net energy producers (transferring energy from its incoming source to the consumers). We therefore recover the essence of the ubiquitous ``food chain'' (or ``food web'') of ecological systems and the ``energy cascade'' of turbulent flow systems.}

\subsection{\label{Proc}{Classification of} Spontaneous Processes}

\begin{table*}[t]
\caption{\label{table_monot}List of possible spontaneous processes in equilibrium and steady state systems, for monotonically varying parameters (terminology from ancient Greek: {\it exo-}, external; {\it endo-}, internal; {\it pan-}, everywhere; {\it tropos}, transformation \citep{Clausius_1865} and {\it -genic}, generating or producing).}
\begin{tabular*}{520pt}{lllccll}
\hline\hline
Case &Conditions &&Entropic Driving Force &Label
&\multicolumn{2}{l}{Extrema at Stationary Position}\\
\hline
\multicolumn{5}{l}{Equilibrium systems ($k d\phi_{eq} = d(G/T) \le 0$)} \\  
E1 &$dS^* \ge 0$,  &$\delta \sigma \ge 0$  &universal &{\it panentropic} &max $S^*$, &max $\sigma$ (= min $H/T$)\\
E2 &$dS^* \ge |\delta \sigma| \ge 0$,  &$\delta \sigma \le 0$ &internal dominant &{\it endoentropic} &max $S^*$, &min $\sigma$ (= max $H/T$)\\
E3 &$dS^* \le 0$,  &$\delta \sigma \ge |dS^*| \ge 0$ &external dominant &{\it exoentropic} &min $S^*$, &max $\sigma$ (= min $H/T$)\\
\hline
\multicolumn{5}{l}{Steady state systems ($K d\phi_{st} \le 0$ with $K=k/{\tscale \Vscale}$)} \\
S1 &$d\mathfrak{H}^*_{st} \ge 0$,   &$\delta \hat{\dot{\sigma}} \ge 0$ &universal &{\it panentropogenic} &max $\mathfrak{H}^*_{st}$, &max $\hat{\dot{\sigma}}$\\
S2 &$K d\mathfrak{H}^*_{st} \ge |\delta \hat{\dot{\sigma}}| \ge 0$,  &$\delta \hat{\dot{\sigma}} \le 0$ &internal dominant &{\it endoentropogenic} &max $\mathfrak{H}^*_{st}$, &min $\hat{\dot{\sigma}}$\\
S3 &$d\mathfrak{H}^*_{st} \le 0$,  &$\delta \hat{\dot{\sigma}} \ge |K d\mathfrak{H}^*_{st}| \ge 0$ &external dominant &{\it exoentropogenic} &min $\mathfrak{H}^*_{st}$, &max $\hat{\dot{\sigma}}$\\
\hline\hline
\end{tabular*}
\end{table*}

\begin{figure*}[t]
\begin{center}
\setlength{\unitlength}{0.6pt}
  \begin{picture}(800,350)
   \put(560,170){\includegraphics[width=50mm]{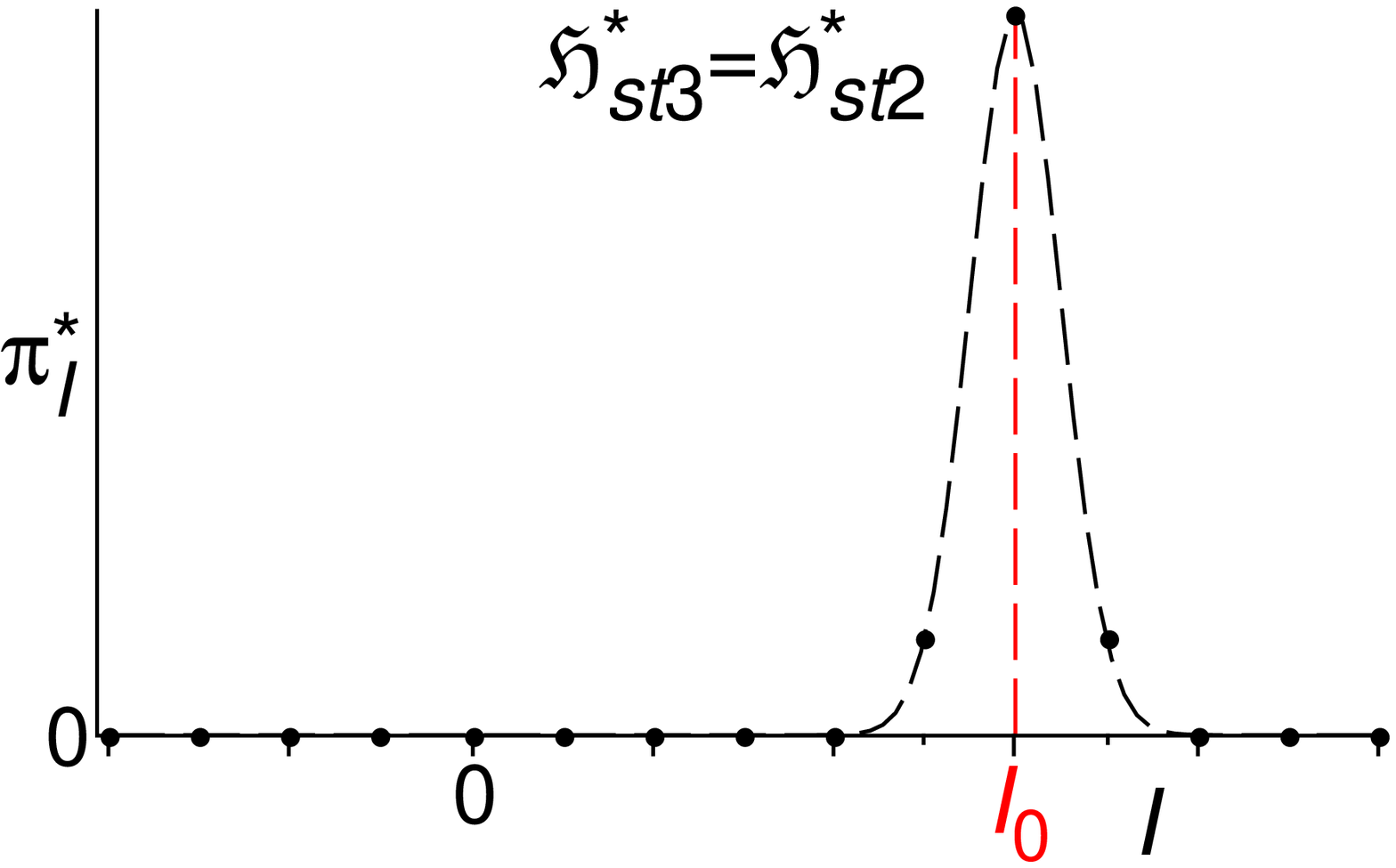} }
   \put(560,160){\small (c)}
   \put(290,170){\includegraphics[width=50mm]{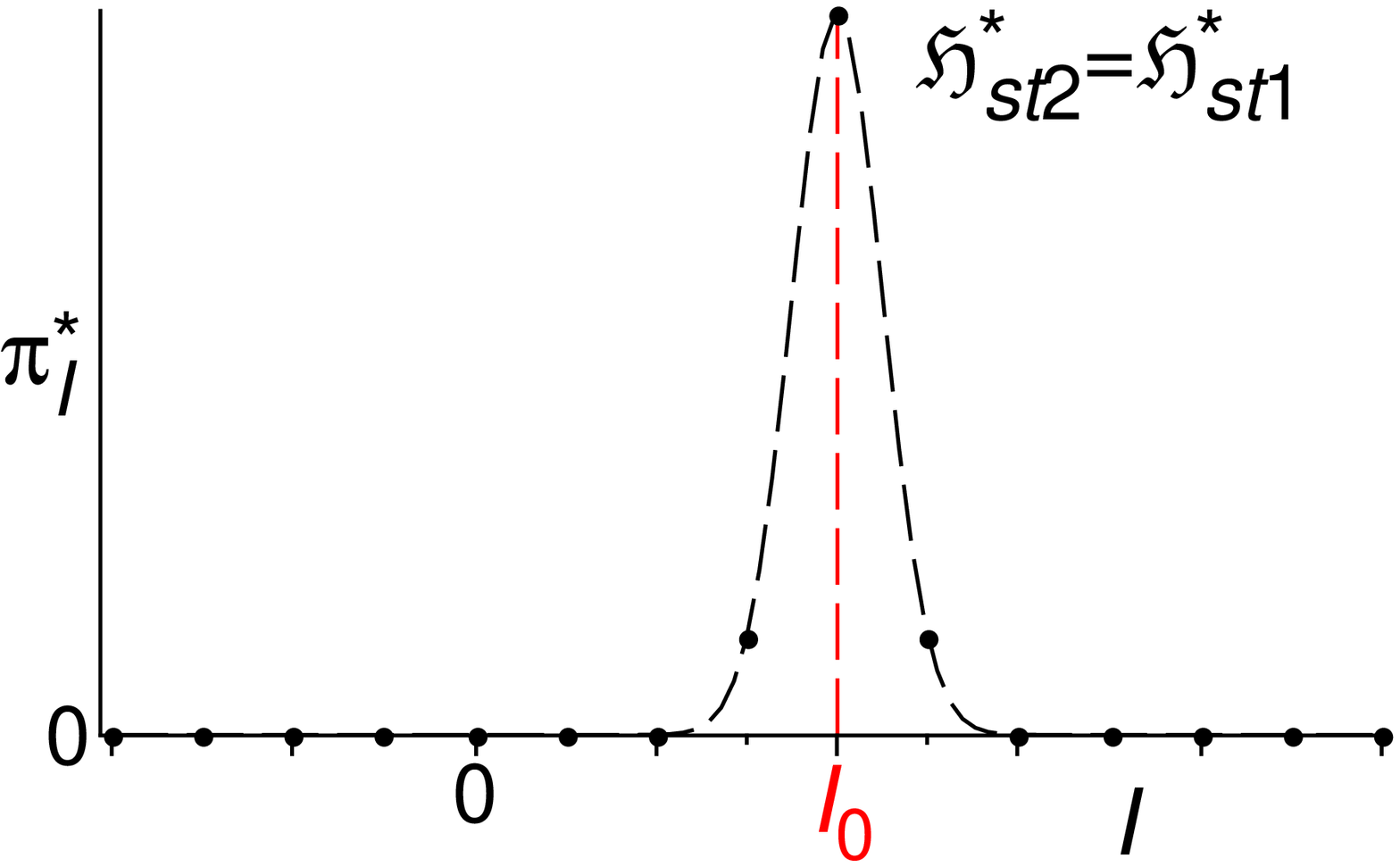} }
   \put(290,160){\small (b)}
   \put(560,10){\includegraphics[width=50mm]{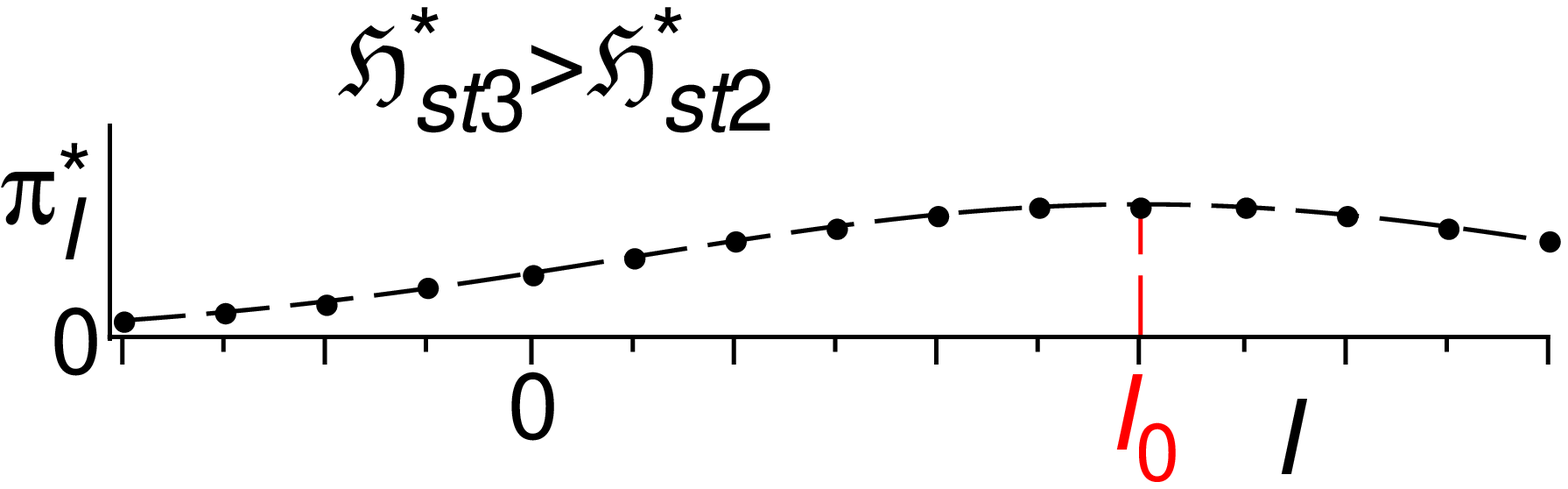} }
   \put(560,0){\small (e)}
   \put(290,10){\includegraphics[width=50mm]{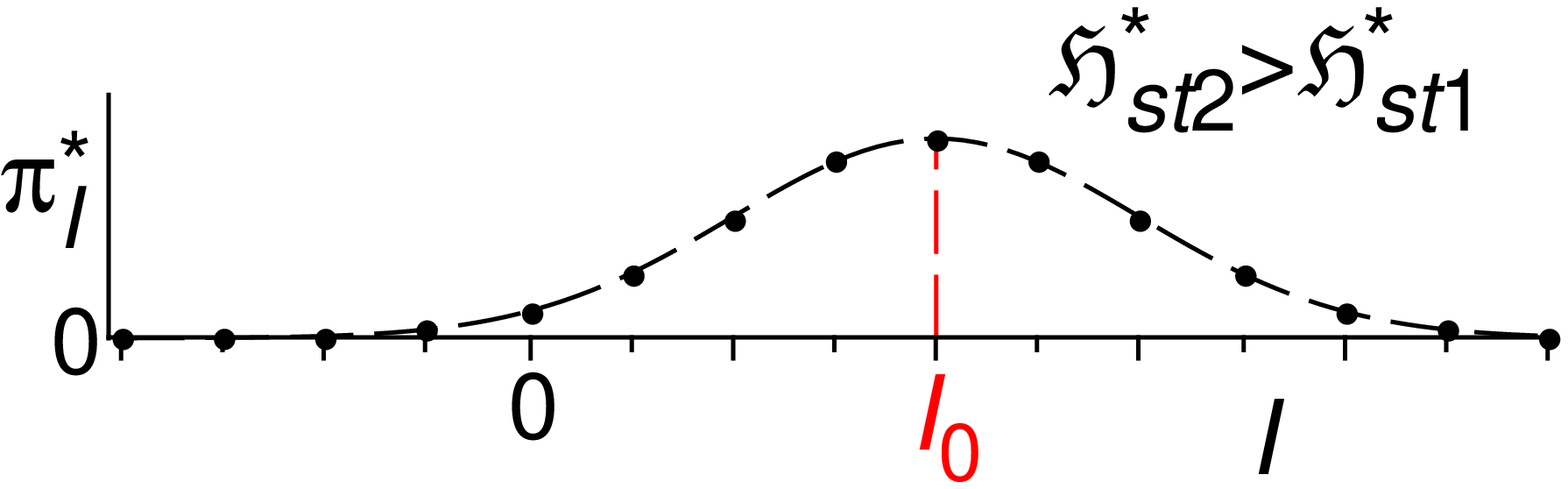} }
   \put(290,0){\small (d)}
  \put(0,60){\includegraphics[width=58mm]{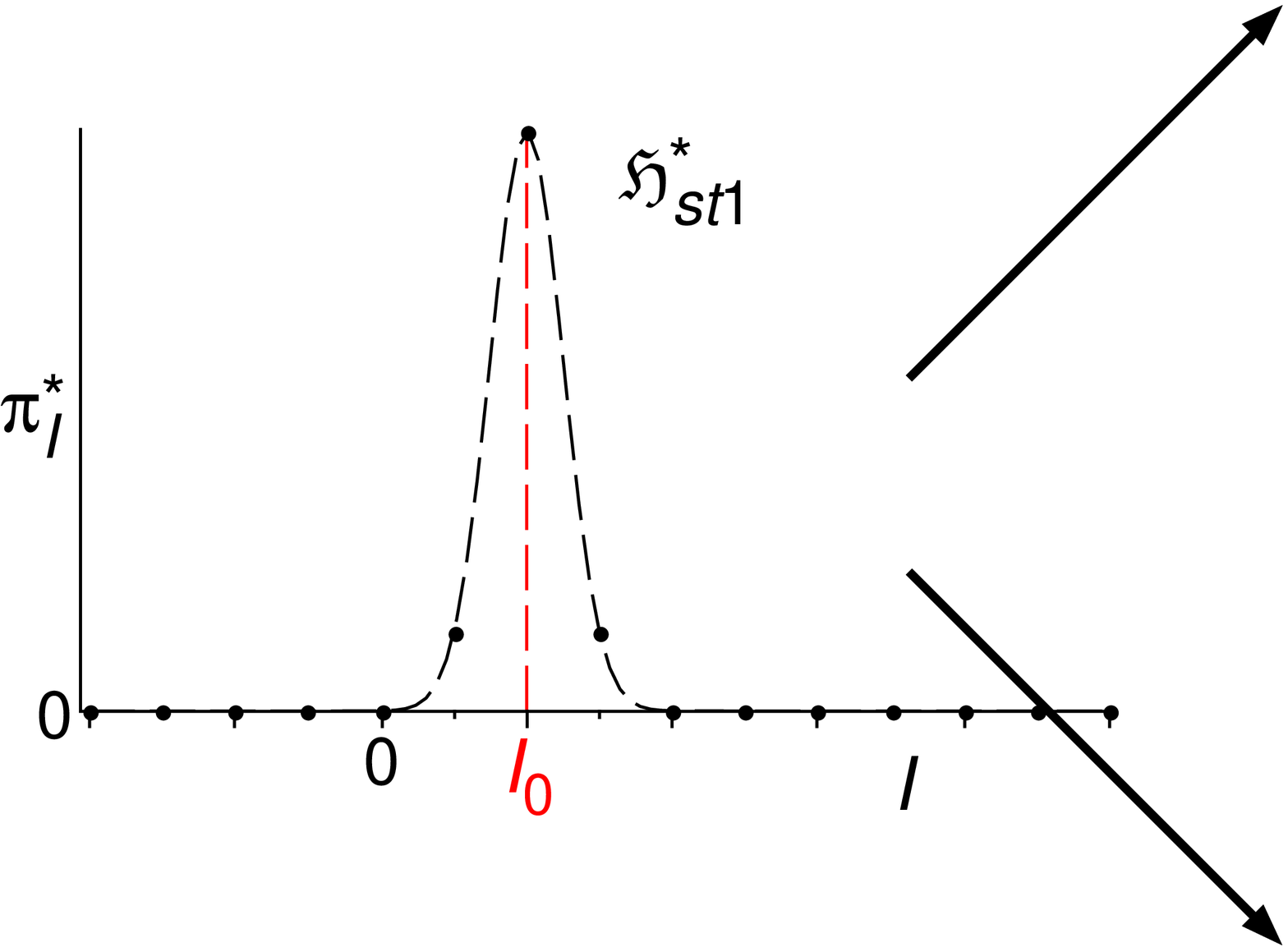} }
   \put(0,50){\small (a)}
  \end{picture}
\end{center}
\caption{Possible responses of a flow system to an increasing force {$F_X$} or mean flux $j_X$ (increasing $I_0$): (a)-(b)-(c) constant $\mathfrak{H}_{st}^*$ or (a)-(d)-(e) increasing $\mathfrak{H}_{st}^*$.} 
\label{fig:h_st2}
\end{figure*}

We can now return to the equilibrium \eqref{eq:Massieu_eq2} and steady state \eqref{eq:Massieu_st} potential functions.  {For simplicity, we first confine the discussion to} processes with monotonic changes in the entropy $S^*$ and entropy produced $\sigma = -H/T$. We see that in equilibrium systems, the path towards equilibrium $d\phi_{eq} \le 0$ will depend on the relative changes in $S^*$ and $\sigma$, {leading to} three possible scenarios for a spontaneous process, as listed in Table \ref{table_monot}.  In Case E1, the process is driven by changes in both entropy terms, whilst in Cases E2 and E3, a reduction in one entropy is ``paid for'' by a greater and opposite gain in the other.  In all cases,  since the entropy changes are monotonic, the equilibrium position (minimum $\phi_{eq}$ = minimum $G/T$) must coincide with extrema (a minimum or maximum) in both $S^*$ and $\sigma$, as set out in the Table.  

Similarly, from \eqref{eq:Massieu_st}, in a flow system subject to monotonic changes in $\mathfrak{H}^*_{st}$ and $\hat{\dot{\sigma}}$, each increment towards steady state $d\phi_{st} \le 0$ could be achieved by one of the three cases listed in Table \ref{table_monot}. The corresponding extrema at steady state are also listed. As shown, Cases S1 and S3 are consistent with a position of {\it maximum entropy production} (MEP). Case S2, on the other hand, involves convergence towards a position of {\it minimum entropy production} (MinEP). The three cases therefore encompass the two major (seemingly contradictory) principles of non-equilibrium thermodynamics {\citep{Prigogine_1967, Prigogine_1980, Martyushev_etal_2007}}. 

We further note that if passage to equilibrium or steady state is not monotonic, many more scenarios are possible.  In thermodynamics, this is handled by considering only the net change 
in Gibbs free energy $\Delta G = -T \Delta S^* + \Delta H$ at constant $T$ and $P$.
In light of \eqref{eq:Massieu_eq2}, this is more {appropriately} written as:
\begin{gather}
k \Delta \phi_{eq} = \Delta \Bigl( \frac{G}{T} \Bigr) = -\Delta S^*  +  \Delta \Bigl( \frac{H}{T} \Bigr) \le 0
\label{eq:deltaMassieu_eq}
\end{gather}
This rests on the fact that $G$, $S^*$, $U$, $V$ and $H$ are state functions, so we can disregard the path taken by the system. Although such systems could follow any of the paths E1-E3 in Table \ref{table_monot} during different stages of the process {\textendash } or even temporarily deviate from $d (G/T) \le 0$ to overcome an activation energy barrier {\textendash } they must approach a position of minimum $G/T$, leading to a net change $\Delta (G/T) \le 0$. The system can still be said to follow one of Cases E1-E3, but now only in a net sense (using $\Delta$'s rather than $d$'s). In Case E3, for example, we can still speak of the system tending towards a position of minimum $S^*$ and maximum $\sigma$ (= minimum $H/T$), provided this is understood to refer to their net changes 
rather than the path taken by the system.

In a similar vein, if an unsteady flow system is not restricted to purely monotonic changes, it must still approach a steady state position of minimum $\mathfrak{H}_{st}^*$, and thus undergo the net change $\Delta \mathfrak{H}_{st}^* \le 0$.  Presuming that $\mathfrak{H}_{st}^*$ and $\hat{\dot{\sigma}}$ can be considered as state functions, the system can still be identified as following {\textendash } now in a net sense {\textendash } one of the three Cases S1-S3 in Table \ref{table_monot}.

Can we infer anything more about flow systems?  Indeed, we can.  Consider a flow element which experiences a gradual increase in the local thermodynamic force {$F_X$} conjugate to the mean local flux or rate $j_X$. Such an element may undergo two types of changes: (i) an increase in $j_X$ without any corresponding increase (or even a decrease) in $\mathfrak{H}_{st}^*$, illustrated in Figures \ref{fig:h_st2}a-b-c; and (ii) increases in both $j_X$ and $\mathfrak{H}_{st}^*$, illustrated in Figures \ref{fig:h_st2}a-d-e. From \eqref{eq:sigma_dot_hat2}, both scenarios involve identical increases in the entropy production, {$\Delta \hat{\dot{\sigma}} = \Delta( j_X F_X)$}. Which is more likely?  From our knowledge of flow systems, the first scenario seems less credible, since it requires the fluxes to remain within a narrow range of instantaneous values at all times, even though the driving force has increased. A decrease in $\mathfrak{H}_{st}^*$ seems even more unlikely. The second scenario permits greater variability (fluctuations) of the fluxes, consistent with the formation of a non-linear mechanism to enable greater transport or production of $X$.  A similar argument applies if the flux, rather than the gradient, is the control variable.  Although this is not a proof, it does lend support to the argument \citep{Niven_MEP} that fluid elements tend to undergo concurrent increases in $\mathfrak{H}_{st}^*$ and $\hat{\dot{\sigma}}$, and thus converge to steady state by a (net) {\it panentropogenic} process. In such cases, the steady state position can be determined by the (net) MEP principle, without concern over contrary effects due to decreases in $\mathfrak{H}_{st}^*$. 

\section{\label{Heur}The MEP ``Heuristic"}
We now turn to a discussion of current practice in the application of the MEP principle to flow or chemical reaction systems, including biological systems. From the pioneering works of \citet{Paltridge_1975, Paltridge_1978} and three decades of further experience \citep[e.g.][]{Ozawa_etal_2001, Ozawa_etal_2003, Juretic_Z_2003, Kleidon_2004, Kleidon_L_book_2005, Dewar_etal_2006, Martyushev_S_2006, Bruers_2007c, Meysman_B_2007}, this has evolved into a set of practices which can be termed the ``MEP heuristic'':
\newcounter{Lcount}
\begin{list}{(\roman{Lcount})}{\usecounter{Lcount} \topsep 3pt \itemsep 3pt \parsep 0pt \leftmargin 22pt \rightmargin 0pt \listparindent 0pt \itemindent 0pt}
\item Divide the control volume into very large subdomains (or even consider the entire domain);
\item Set up the set of mass, chemical species, energy, momentum and/or charge balance equations for the system, based on the bulk flow rates between subdomains, using linear (Onsager-like) transport equations with adjustable, whole-subdomain transport coefficient(s), and chemical reaction rate equations with adjustable first-order rate constant(s);
\item Calculate the thermodynamic entropy production of the system, as a function of the adjustable parameter(s);
\item The inferred steady state of the system is given by the position of maximum entropy production with respect to the adjustable parameter(s).
\end{list}

How does this heuristic work?  In effect, it selects the highest allowable entropy production consistent with the set of allowable bulk net fluxes $J_{X,\Gamma}$ and bulk thermodynamic forces $F_{X,\Gamma}$ in and between subdomains $\Gamma$ of the system:
\begin{equation}
\text{MEP\_ Heuristic} = \max\limits_{\vect{\Omega}} \; \Biggl( \sum\limits_{\Gamma}  \sum\limits_X J_{X,\Gamma}(\vect{\Omega}) \; F_{X,\Gamma}(\vect{\Omega}) \Biggr)
\label{eq:MEPH}
\end{equation}
where the bulk fluxes and/or forces are functions of the set of subdomain-wide adjustable parameters $\vect{\Omega} = \{\Omega\}$.  It must be recognised, however, that the adjustable parameters are secondary variables, which do not represent fundamental physical processes. The true maximum must therefore be given by a ``system maximum entropy production'' (SMEP) principle:
\begin{equation}
\text{SMEP} = \max \Biggl( \iiint\limits_{CV} \hat{\dot{\sigma}}(V) \, dV \Biggr) 
\label{eq:SMEP}
\end{equation}
where the maximum is taken with respect to the instantaneous fluxes $j_{X,\vecti}$, conditioned by the constraints on the system, and the integral is calculated over the control volume.  The MEP heuristic therefore makes the {assumption} that \eqref{eq:MEPH} and \eqref{eq:SMEP} are equivalent, which is correct if and only if there exists a set of local physical mechanisms by which the maximum in \eqref{eq:MEPH} can be physically realised.  Using the terminology of MEP practitioners, the MEP heuristic \eqref{eq:MEPH} must be considered to apply only to ``many-degree-of-freedom'' systems
\citep[e.g.][]{Ozawa_etal_2001, Ozawa_etal_2003, Juretic_Z_2003, Kleidon_2004, Kleidon_L_book_2005, Dewar_etal_2006, Martyushev_S_2006, Bruers_2007c, Meysman_B_2007}.

In contrast, the analysis herein (\S\ref{Gen}-\ref{Imp}) and in \citet{Niven_MEP} gives the optimisation principle:
\begin{equation}
\begin{split}
\text{Optimum} &=\iiint\limits_{CV}  \min \bigl( \phi_{st} (V) \bigr) \, dV  \\
&= \iiint\limits_{CV} \max \Bigl( \mathfrak{H}_{st}^*(V) + \frac{\theta \flow{V} \, \hat{\dot{\sigma}} (V)   }{k} \Bigr) \, dV 
\label{eq:Opt}
\end{split}
\end{equation}
If the parameters $\mathfrak{H}_{st}^*$ and $\hat{\dot{\sigma}}$ are positively correlated {\textendash } as argued in \S\ref{Imp} {\textendash } then \eqref{eq:Opt} becomes functionally equivalent to a ``local maximum entropy production'' (LMEP) principle, which gives for the overall system:
\begin{equation}
\text{LMEP} =  \iiint\limits_{CV} \bigl( \max \hat{\dot{\sigma}} (V) \bigr) \, dV 
\label{eq:LMEP}
\end{equation}
This is a much stronger condition than \eqref{eq:SMEP}.  By considerations of integral calculus \citep{Zwillinger_2003}, the two bounds are related by:
\begin{equation}
\max \Biggl( \iiint\limits_{CV} \hat{\dot{\sigma}}(V) \, dV \Biggr)
\le  \iiint\limits_{CV}  \max  \hat{\dot{\sigma}}(V) \, dV
\label{eq:LMEP_SMEP}
\end{equation}
since the left hand side could possess regions of $\hat{\dot{\sigma}} < 0$, compensated by other regions of greater $\hat{\dot{\sigma}} > 0$.  This, however, runs against an argument used by \citet{Prigogine_1967, Prigogine_1980}: how can a system possibly ``know'' that it can consume entropy in some regions, which will be compensated by greater entropy production in others? Indeed, we could construct a smaller control volume containing only the entropy-consuming elements, which would continuously violate the second law of thermodynamics. It is for this sound reason that the MEP principle must be a local principle, applicable at all volume scales.  With the restriction $\hat{\dot{\sigma}}(V) \ge 0$, we see that the two maxima in \eqref{eq:LMEP_SMEP} coincide, and so the MEP heuristic (with its assumption of many degrees of freedom) becomes equivalent to the local formulation.

%
\section{\label{Concl}Conclusions}
This study examines the meaning and implications of a new formulation of non-equilibrium thermodynamics applicable to  flow and/or chemical reaction systems at steady state \citep{Niven_MEP}. This provides a very different, conditional derivation of the ``maximum entropy production'' (MEP) principle, based on minimisation of a dimensionless, local, free-energy-like potential function $\phi_{st}$. The analysis encompasses all biological and ecological systems. Firstly, the basis of the derivation and the meaning of $\phi_{st}$ are examined. The flux entropy $\mathfrak{H}_{st}^*$ used in the analysis is then shown to represent the ``spread'' of the distribution of instantaneous fluxes and/or reaction rates through or within the element. Since a flow system can access states of reverse flow or reaction, a high flux entropy is consistent with {\it higher} {\it variability} and thus with {\it chaotic} or {\it oscillatory} processes. In this respect, the term ``steady state'' is therefore something of a misnomer, since it refers only to the constancy of the mean bulk flows and not their temporal and spatial variability.  {One consequence, examined through a specific example, is the coexistence of energy producers and consumers in ecological systems.} 

The effects of reinforcement or competition between changes in flux entropy $\mathfrak{H}_{st}^*$ and entropy production $\hat{\dot{\sigma}}$ are then examined and classified. It is argued that in many systems, these two parameters should increase concurrently, enabling the steady state position to be determined by the MEP principle.  The ``MEP heuristic" used by MEP practitioners is then shown to be consistent with the present local formulation, with the additional assumption that the system has sufficient dynamic degrees of freedom that the MEP state can be physically realised.

\section*{Acknowledgments}
The author thanks the participants of the MEP workshops hosted by the Max-Planck-Institut f\"ur Biogeochemie, Jena, Germany, in 2007 and 2008, for valuable discussions; The University of New South Wales and the above Institute for financial support; and the European Commission for financial support as a Marie Curie Incoming International Fellow (2007-2008) under Framework Programme 6.


\end{document}